\documentclass[10pt,journal,compsoc]{IEEEtran}

\ifCLASSOPTIONcompsoc
  \usepackage[nocompress]{cite}
\else
  \usepackage{cite}
\fi
\ifCLASSINFOpdf
  \usepackage[pdftex]{graphicx}
\else
\fi
\usepackage{balance}

\usepackage{url}
\usepackage{hyperref}

\usepackage{mfirstuc}

\usepackage{amsmath}
\usepackage{paralist}
\usepackage{multirow}
\usepackage{makecell}
\usepackage{caption}
\usepackage{algorithm} 
\usepackage{algorithmic}  
\usepackage[algo2e]{algorithm2e} 

\usepackage{dblfloatfix}
\usepackage{balance}
\usepackage[dvipsnames]{xcolor}

\newcommand{\gesLen}{{\textit{GestureLens}}}

\newcommand{\gesLenVideoview}{{video view}}
\newcommand{\gesLenVideoviewCapital}{{Video View}}
\newcommand{\gesLenContentview}{{relation view}}
\newcommand{\gesLenContentviewCapital}{{Relation View}}

\newcommand{\gesLenMatrixview}{{exploration view}}
\newcommand{\gesLenMatrixviewCapital}{{Exploration View}}

\newcommand{\gesLenDynamicview}{{dynamic view}}
\newcommand{\gesLenDynamicviewCapital}{{Dynamic View}}

\usepackage{tikz}

\newcommand\copyrighttext{%
  \footnotesize \textcopyright 2022 IEEE. Personal use of this material is permitted.
  Permission from IEEE must be obtained for all other uses, in any current or future
  media, including reprinting/republishing this material for advertising or promotional
  purposes, creating new collective works, for resale or redistribution to servers or
  lists, or reuse of any copyrighted component of this work in other works.
  DOI: \href{10.1109/TVCG.2022.3169175}{10.1109/TVCG.2022.3169175}}
\newcommand\copyrightnotice{%
\begin{tikzpicture}[remember picture,overlay]
\node[anchor=south,yshift=10pt] at (current page.south) {\fbox{\parbox{\dimexpr\textwidth-\fboxsep-\fboxrule\relax}{\copyrighttext}}};
\end{tikzpicture}%
}

\begin{document}
%
\title{\gesLen: Visual Analysis of Gestures in Presentation Videos}
%
%
%
%

\author{Haipeng Zeng, Xingbo Wang, Yong Wang, Aoyu Wu, Ting Chuen Pong and Huamin Qu,~\IEEEmembership{Member,~IEEE,}
\IEEEcompsocitemizethanks{
\IEEEcompsocthanksitem H. Zeng is with Sun Yat-sen University.\protect\\
E-mail: zenghp5@mail.sysu.edu.cn

\IEEEcompsocthanksitem X. Wang, A. Wu, T. Pong and H. Qu are with the Hong Kong University of Science and Technology.\protect\\
E-mail: \{xwangeg, awuac, tcpong, huamin\}@cse.ust.hk

\IEEEcompsocthanksitem Y. Wang is with the Singapore Management University.\protect\\
E-mail: yongwang@smu.edu.sg
}
\thanks{Manuscript received June 19, 2021; revised April 19, 2022.}}

%
%

\markboth{IEEE TRANSACTIONS ON VISUALIZATION AND COMPUTER GRAPHICS}%
{Shell \MakeLowercase{\textit{et al.}}: Bare Advanced Demo of IEEEtran.cls for IEEE Computer Society Journals}
%



\IEEEtitleabstractindextext{%
\begin{abstract}
Appropriate gestures can enhance message delivery and audience engagement in both daily communication and public presentations.
In this paper, 
we contribute a visual analytic approach that assists professional public speaking coaches in improving their practice of gesture training through analyzing presentation videos.
Manually checking and exploring gesture usage in the presentation videos is often tedious and time-consuming. 
There lacks an efficient method to help users conduct gesture exploration,
which is challenging due to the intrinsically temporal evolution of gestures and their complex correlation to speech content. 
In this paper, we propose {\gesLen}, a visual analytics system to facilitate gesture-based and content-based exploration of gesture usage in presentation videos.
Specifically,
the {\gesLenMatrixview} enables users to obtain a quick overview of the spatial and temporal distributions of gestures.
The dynamic hand movements are firstly aggregated through a heatmap in the gesture space for uncovering spatial patterns, 
and then decomposed into two mutually perpendicular timelines for revealing temporal patterns.
\textcolor{black}{The {\gesLenContentview} allows users to explicitly explore the correlation between speech content and gestures by enabling linked analysis and intuitive glyph designs.}
\textcolor{black}{
The {\gesLenVideoview} and {\gesLenDynamicview}
show the context and overall dynamic movement of the selected gestures, respectively.}
\textcolor{black}{Two usage scenarios and expert interviews with professional presentation coaches demonstrate the effectiveness and usefulness of {\gesLen} in facilitating gesture exploration and analysis of presentation videos.}
\end{abstract}

}

\maketitle
\copyrightnotice

\IEEEdisplaynontitleabstractindextext

%
\IEEEpeerreviewmaketitle

\vspace{-3mm}

\ifCLASSOPTIONcompsoc
\IEEEraisesectionheading{\section{Introduction}\label{sec:introduction}}
\else
\section{Introduction}
\label{sec:introduction}
\fi

%
%
%
%

\IEEEPARstart{G}{estures} play an important role in communication and presentations~\cite{kendon2004gesture, kendon1994gestures}.
Speakers usually use gestures spontaneously when producing utterances~\cite{tieu2017co} and giving explanations~\cite{kang2016hands, alibali1999illuminating}. 
Many studies have emphasized the importance of coherence in gesture and verbal content in public speaking when delivering messages~\cite{lucas2009art, hoogterp2014your, Toastmasters2011Gestures}. 
For example, hand gestures should convey the same meaning as the verbal content while avoiding distracting audiences~\cite{Toastmasters2011Gestures}.
Therefore, appropriate gestures have a great influence on message delivery and audience engagement. 

However, it is challenging to use appropriate gestures in daily communications and public presentations.
Existing guidelines for gestures are mainly theoretically derived,
leading to misalignment with practical and theoretical sources~\cite{carstens2019advice}.
Some guidelines even occasionally contradict others, which can confuse speakers~\cite{wu2018multimodal}. 
For example, presentation expert Khoury~\cite{Khoury2017five} encourages speakers to use more hand movements in presentations, 
while Currie~\cite{Currie2015ten} criticizes continuous hand movements.
Therefore, 
it is hard for presentation coaches to train speakers on gesture usage.

In this paper, we closely work with coaches from a professional presentation training company to improve their practice of gesture training.
Analyzing presentation videos is one of the common methods they use on training gesture usage.
Usually, it is hard for speakers to get aware of the performance of their own gestures. 
Coaches analyze good presentation videos or videos recording speakers' practices, 
which can provide some examples and evidence for improvement. 
\textcolor{black}{Thus, analyzing presentation videos is of great value to understand gesture usage.
However, manually checking and exploring gesture usage in the presentation videos is often tedious and time-consuming.
In this paper,
we aim to help coaches explore and analyze gesture usage in presentation videos.
Particularly, as a first step, we focus on hand movements because of their great contributions to the body language in presentations~\cite{beattie2016rethinking, haider2016presentation}.}

Some existing tools~\cite{wittenburg2006elan, kipp2014anvil} have been proposed to facilitate analyzing gestures in collected videos. 
For example, ELAN~\cite{wittenburg2006elan} helps annotates gestures in videos. 
But, it is time-consuming to watch videos one by one, let alone to explore and analyze gestures in presentation videos. 
Other studies~\cite{wan2016chalearn, materzynska2019jester, poppe2010survey, wu2016deep} have focused on automatically recognizing gestures. 
They only focus on limited types of gestures without regard to verbal content. 
There are some attempts~\cite{jang2014gestureanalyzer, bernard2013motionexplorer, jang2015motionflow} to adopt visualization techniques to analyze collected motion data, which allows users to explore motion data by leveraging their own knowledge. 
However, these methods mainly visualize and explore motion data without considering speech content in presentation scenarios.
Therefore, to better analyze gesture usage in presentation videos, an interactive visualization system that supports exploring gesture and its complex correlation with speech content would be highly valuable for users.

Visually analyzing the gestures from presentation videos is a nontrivial task due to three major reasons:
\textbf{(1) Dynamic process}. 
Gestures in presentation videos can be regarded as high dimensional time-series data, since it includes the movement of multiple joints.
The stochastic and dynamic nature process of human motion brings significant challenges in measuring and understanding gestures in presentation videos.
\textbf{(2) \textcolor{black}{Complex gesture categories}}. 
Different people have different gesture styles in presentation scenarios. 
Most work mainly focuses on a few categories of gestures, which is not sufficient in fully representing various gestures used in presentation scenarios. 
It is difficult to summarize and digest different gestures without clearly defined categories.
\textbf{(3) Hidden relationships}. People usually produce spoken utterances accompanying a series of gestures spontaneously. 
The meaning of gestures is related to various kinds of speech content. Moreover, gestures sometimes have no apparent meaning~\cite{alibali2001effects}. Gestures are often related to speech content. Thus, without speech content, it is hard to interpret gestures used in different presentations.


To address the aforementioned challenges, we developed an interactive visualization system to support the exploration of gestures and the relationship between gestures and speech in presentation videos.
We worked closely with experienced public speaking coaches from the international training company for about six months to derive the analytical tasks. 
Based on the derived tasks, 
four coordinated views are designed and implemented to support two basic exploration schemes (i.e. gesture-based and content-based exploration) on gesture analysis in presentation videos.
Specifically, to describe dynamic hand movements, a heatmap in the gesture space and two mutually perpendicular timelines in the {\gesLenMatrixview} are designed to visualize the spatial and temporal distributions of gestures. 
\textcolor{black}{To better analyze different gestures, we mainly adopt cluster algorithms to group similar gestures together. Users are allowed to leverage their prior knowledge for gesture analysis. The bottom part of the {\gesLenContentview} shows the similarities between different gestures.
To reveal the hidden relationships between gestures and speech content, human stick figures are shown on the top of each word of the transcript in the {\gesLenMatrixview}.
Further, users are allowed to explore the correlation between gestures and speech content in the {\gesLenContentview} with the linked graph design.
In addition, the {\gesLenVideoview} and {\gesLenDynamicview} provide users with further details on the video analysis.}
Rich interactions are provided to enhance our system.
\textcolor{black}{Finally, two usage scenarios and expert interviews demonstrate the effectiveness and usefulness of our system.}



In summary, our primary contributions are as follows:
\begin{itemize}
\item We propose an interactive visual system, {\gesLen}, to facilitate gesture analysis in presentation videos from multiple aspects, such as spatial and temporal distributions of gestures and their correlation with the speech content.


\item We propose effective visual designs to enable interactive exploration of gestures. Particularly, enhanced horizontal and vertical timelines reveal the temporal distribution of gestures, and the linked graph with human stick-figure glyphs explicitly uncovers the correlation between gestures and speech content.

\item We present two usage scenarios and conduct expert interviews with professional presentation coaches, which provides support for the usefulness and effectiveness of {\gesLen} in exploring gesture usage and speech content of presentation videos.
\end{itemize}


\section{Related Work}
This section presents three relevant topics, the correlation between gestures and speech content, human motion analysis, and visual analysis of presentation techniques.

\vspace{-1mm}
\subsection{Correlation between gestures and speech content}
Gestures mainly refer to the hand movements that facilitate message delivery~\cite{kendon2004gesture}. 
Much research has been conducted to analyze gestures in communication.
For example, researchers have defined the gesture space~\cite{mcneill1992hand, gunter2015inconsistent} to describe the spatial characteristics and have segmented gestures into different phases to reveal the temporal patterns of gestures~\cite{kendon2004gesture, bressem2011rethinking}.
It is widely believed that there exist correlations between gestures and speech content~\cite{wagner2014gesture}. 
Researchers identify six types of spontaneous gestures that are combined with speech~\cite{carstens2019advice}, i.e., iconic, metaphoric, deictic, beats, emblems and discourse.

Prior studies have analyzed the relationship between gestures and speech by manually labeling the motion data. 
For example, Denizci and Azaoui~\cite{denizci2015reconsidering} manually analyzed how teachers use the gesture space to convey meaning properly in classroom settings and Bressem and Ladewig~\cite{bressem2011rethinking} explored gesture phases with articulatory features.
Automated methods have been developed to facilitate gesture exploration.
For example, Madeo et al.~\cite{madeo2016gesture} segmented the gesture phase with support vector machine algorithms. 
Further, Okada et al.~\cite{okada2013context} proposed a framework for classifying communicative gestures with the contextual features from narrative speech.
These studies shed light on the capturing relationships between gestures and speech content. 
Inspired by them, some recent research has directly modeled the relationship between gestures and speech content. 
For example, Yoon et al.~\cite{yoon2019robots} proposed an end-to-end neural network model trained on TED talks to generate sequential gestures from the input text.
Ginosar et al.~\cite{ginosar2019learning} studied the connection between speech and gestures and further proposed a model to generate gestures from audio.

In this paper, 
we enhance automated methods with novel visualization techniques to help users analyze gestures with their domain knowledge.

\vspace{-1mm}
\subsection{Human Motion Analysis}
Human motion analysis has attracted much attention from researchers. Both automated methods and visualization techniques have been applied to analyze human motion.

Some motion labeling tools~\cite{wittenburg2006elan, lausberg2009coding, dael2012body} are proposed to allow the creation of gesture annotations in videos. However, such kinds of tools are very time-consuming and labor-intensive. 
Some researchers have turned to automatic methods, while other recent studies~\cite{wan2016chalearn, materzynska2019jester, poppe2010survey, wu2016deep} have focused on automatic human motion detection and recognition. 
Many works focus on unsupervised methods~\cite{zhou2012hierarchical, bernard2013motionexplorer,jang2014gestureanalyzer,jang2015motionflow}. 
For example, Zhou et al.~\cite{zhou2012hierarchical} proposed an unsupervised hierarchical aligned cluster analysis algorithm to cluster human motion, which shows the power of clustering methods in analyzing gesture data.

Visualization is an intuitive and effective way for exploring motion data. Some visualization approaches have been proposed to facilitate people in exploring gesture data. A detailed survey is conducted by Bernard et al.~\cite{bernard2017approaches}.
For example, Hilliard et al.~\cite{hilliard2017technique} visualized the physical trajectory of movement using video. 
Ginosar et al.~\cite{ginosar2019learning} combined heatmap and human stick figure to show the overview and movement of hand gestures. 
Further, Bernard et al.~\cite{bernard2013motionexplorer} presented MotionExplorer to visually explore human motion data on hierarchical aggregation. 
Similarly, Jang et al.~\cite{jang2015motionflow} proposed MotionFlow to visually abstract and aggregate sequential patterns in human motion data. 
However, these studies cannot be directly applied to our scenarios since they do not consider the semantic meaning of gestures and the relationship between gestures and speech content.
 
In this paper, we target presentation scenarios and propose a visual analytics system to facilitate the exploration of gesture usage as well as the correlation with speech content. 

\vspace{-1mm}
\subsection{Visual Analysis of Presentation Techniques}

Visualization approaches have been widely used to analyze presentation techniques from different dimensions, such as text, audio, facial expression.
For example, Tanveer et al.~\cite{tanveer2018awe} analyzed the narrative trajectories using the transcripts of over 2000 TED talks, which reveals the relationship between narrative trajectories and the ratings by the audience. 
Xia et al.~\cite{xia2022persua} developed~\textit{Persua} which offers example-based guidance to improve persuasiveness of arguments in online discussion.
Yuan et al.~\cite{yuan2019speechlens} introduced~\textit{SpeechLens} to explore and identify narration strategies in public speaking by analyzing textural and acoustic information.
Wang et al.~\cite{wang2020voicecoach} presented ~\textit{VoiceCoach} to explore voice modulation skills in 2,623 TED talks and facilitate effective training on voice modulation skills.
Wang et al.~\cite{wang2021dehumor} designed \textit{DeHumor} to explore verbal content and vocal delivery of humor snippets in public speaking.
These studies mainly focus on the verbal aspects of presentation techniques.
Different from them, other research papers~\cite{zeng2019emoco,wang2021m2lens,zeng2020emotioncues} have also investigated facial expressions for non-verbal communication. For example, Zeng et al.~\cite{zeng2019emoco} proposed~\textit{EmoCo} to analyze emotion coherence between facial expressions, speech content, and speakers' voice in presentation videos.
However, all these studies have not explored gesture usage, one of the most important factors in presentations.

Several visual analytics systems on gesture analysis have also been proposed. 
Tanveer et al.~\cite{tanveer2016automanner} presented AutoManner, an intelligent interface that automatically extracts human gestures from motion capture data, which makes speakers aware of their mannerisms. 
However, it does not explicitly explore the relationship between gesture and semantic content, and fails to interpret meaningful gestures associated with the speech content.
Further, Tanveer et al.~\cite{tanveer2017automatic} proposed a framework to automatically identify non-meaningful body-movements in the context of speech. 
In addition, Wu and Qu~\cite{wu2018multimodal} explicitly explored the relationship between speech content and gestures. 
However, this work mainly provides a high-level and coarse analysis of gesture and speech content. 
In this paper, we focus on analyzing gestures in presentations and provide a fine-grained analysis of gestures and speech content.


\begin{figure*}[!t]
 \centering
 \includegraphics[width=2\columnwidth]{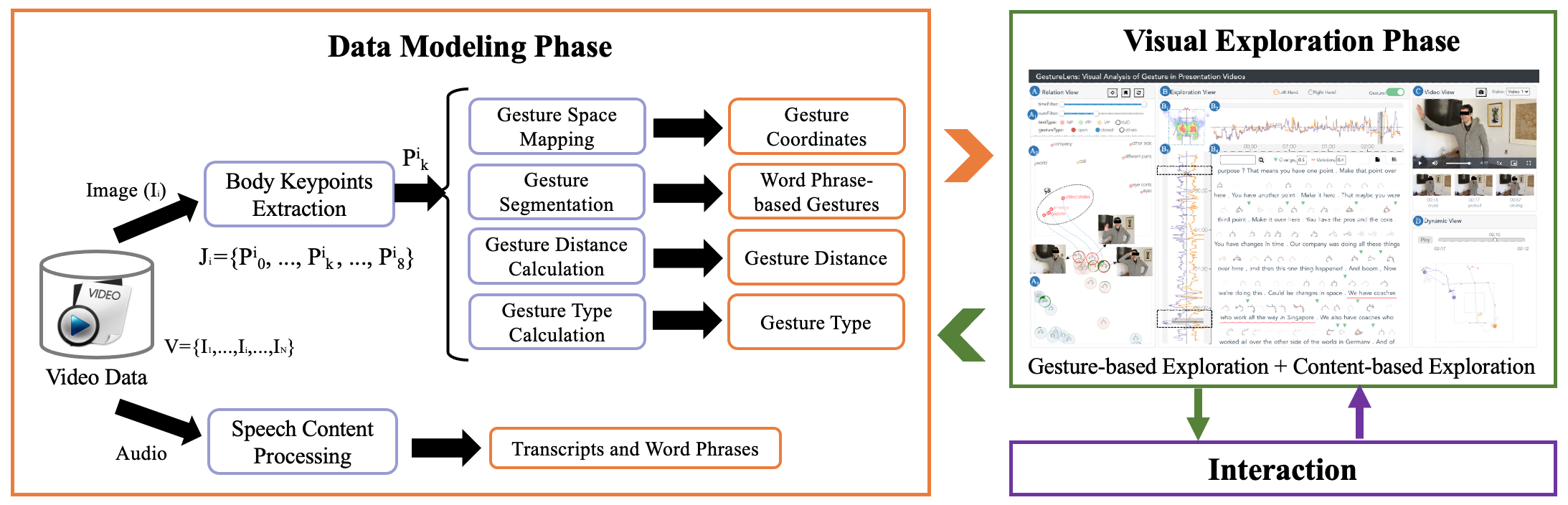}
 \vspace{-2mm}
  \caption{Our visualization system pipeline: in the data modeling phase, we use well-established methods to extract body keypoints from videos and conduct data processing. To be specific, a video is modeled as a series of frames. After detecting the body keypoints of each frame, we obtain joint coordinates. Then we conduct data processing, e.g., gesture segmentation. The transcripts and word phrases are obtained by using speech-to-text techniques and NLP techniques. In the visual exploration phase, four coordinated views are provided to support gesture- and content-based exploration.}
  \label{fig:gesLenPipeline}
\end{figure*}

\section{Data and Analytical Tasks}
In this section, we first describe the data processing procedure and then summarize a set of analytical tasks based on discussions with our domain experts.

\subsection{Data Processing} 
\label{section:gesLenDataProcessing}
In this paper, we focus on those videos recording a speaker's presentation, where the speaker stands in front of the camera.
Given a presentation video, the video can be modeled as a series of images: $V=\{I_1, I_2, ..., I_i, ..., I_N\}$, where $I_i$ indicates $i$-th frame and $N$ indicates the frame number in the video. Then we conduct data processing (Fig.~\ref{fig:gesLenPipeline}) with widely used algorithms and techniques to extract the speaker's gestures and corresponding speech transcripts.
The major steps of data processing are described as follows. 

\begin{figure}[!htb]
  \setlength{\belowcaptionskip}{-10pt}
  \centering
  \includegraphics[width=0.38\columnwidth]{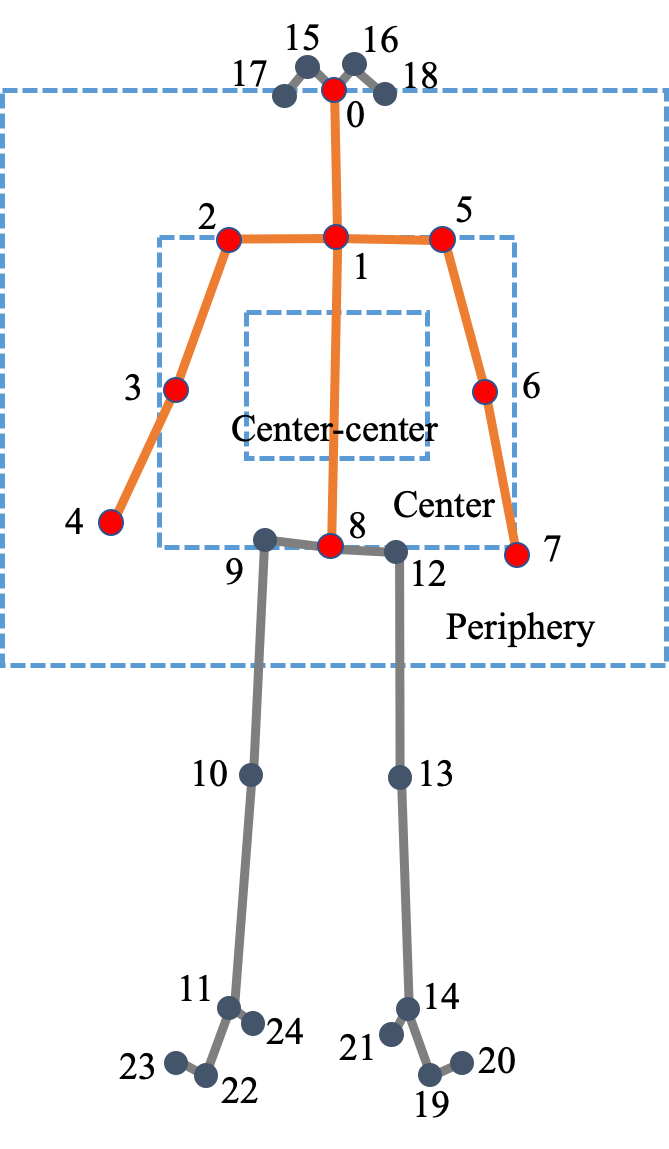}
  \vspace{-4mm}
  \caption{McNeill's gesture space diagram\cite{mcneill1992hand} includes 25 body keypoints. We focus on the upper body (0-8 body keypoints), which is highlighted with red dots and orange lines. Gesture space is described by three blue dashed rectangles.}
  \label{fig:gesLenGestureSpace}
\end{figure}

\textbf{Body Keypoint Detection.}
To obtain the detailed gestures of a speaker, we need to detect body keypoints first. 
Here we adopt OpenPose~\cite{Cao2019openpose}, a widely used real-time multi-person keypoint detection library for body and hand estimation. For each frame $I_i$, we can detect the corresponding body keypoints, $J_i = [P_{0}^{i}, P_{1}^{i}, ... ,P_{k}^{i},..., P_{24}^{i}]$, where $P_{k}^{i}=[x_k, y_k, c_k]$ indicates the coordinates, $x_k$ is x-axis value, $y_k$ is y-axis value and $c_k$ is the confidence probability. Since we mainly focus on the upper body hand gestures, we extract 9 body keypoints of the upper body, which are highlighted in red as shown in Fig.~\ref{fig:gesLenGestureSpace}.

\textbf{Gesture Space Mapping.}
After detecting the body keypoints, we can further obtain gesture coordinates.
However, there are two major issues in analyzing gestures, i.e., \textit{coordinate normalization} and \textit{spatial description of gestures}.
Coordinate normalization is important because it is hard to measure the gesture differences between different frames or presentation videos without a unified space.
Thus, we normalize the gesture coordinates based on Keypoint 0 (the red circle annotated ``0" in Fig.~\ref{fig:gesLenGestureSpace}). 
To be specific, we regard Keypoint 0 as the coordinate origin.
The coordinates of other keypoints are calculated based on Keypoint 0.
Further, we normalize coordinates to $[-1\sim1]$ range by using the height of the person. 
For the spatial description of gestures, we employ the gesture space defined in McNeill's gesture space theory~\cite{mcneill1992hand}, as shown in the blue dashed rectangles of Fig.~\ref{fig:gesLenGestureSpace}. 
The three-level dashed rectangles represent the spatial regions of the \textit{center-center}, \textit{center}, and \textit{periphery} from the inner to the outside. 
Specifically, the \textit{center-center region} is the area directly in front of the chest; 
the \textit{center region} is the area surrounding the \textit{center-center region}, which stretches from the shoulders down to the waist and covers both sides of the body; 
the \textit{periphery region} stretches from ears to the knees, surrounding the \textit{center-center region}.

\textcolor{black}{
\textbf{Gesture Segmentation.}
A speaker often uses various gestures for different speech content. 
We segment gestures according to word phrases, 
since we are interested in the relationships between gestures and speech content.}

\textbf{Gesture Distance Calculation.}
To find similar gestures, it is necessary to define gesture distance between different continuous gestures.
Since gestures may contain different frame lengths,
we first define a distance function for measuring similarities between two static frames; 
then we utilize dynamic time warping (DTW)~\cite{muller2007dynamic}, a distance measure algorithm for time-series data with variable lengths.
Given two keypoint coordinates in frames $F$ and $G$, and each keypoint vector is represented as $P_{k}^{i}=[x_k, y_k, c_k]$, where $k$ indicates $k$-th keypoint in $F$ and $G$, $x_k$ is x-axis value, $y_k$ is y-axis value and $c_k$ is the confidence probability, the distance between gesture in two frames can be defined as~\cite{Friedhoff2018Move}:
$$ D(F, G) = \frac{1}{\sum_{k=0}^{8} F_{c_k}} * \sum_{k=0}^{8} F_{c_k} * \|F_{xy_k}\ - G_{xy_k}\| $$
where $F_{c_k}$ is the confidence probability for $k$-th keypoint, $F_{xy_k}$ and $G_{xy_k}$ is the coordinates of $k$-th keypoint of $F$ and $G$, respectively. 

\textcolor{black}{
\textbf{Gesture Type Calculation.}
After collecting different word phrase-based gestures, 
we classify gestures into three different categories, i.e., \textit{closed gestures}, \textit{open gestures}, and \textit{others}~\cite{wu2018multimodal}. \textit{Closed gestures} refer to gestures where hands are put closely or overlapped with the torso; 
\textit{Open gestures} refer to gestures where two hands are far away from each other and wrist points go outermost. For those gestures where hands are fall in the torso region, we named them \textit{others}. More types can be calculated based on users' requirements.
}

\textcolor{black}{
\textbf{Speech Content Processing.}
\textcolor{black}{For a presentation video, the transcript can be obtained by adopting automatic transcription techniques\footnote{https://aws.amazon.com/transcribe/}.
Further, to support word phrase analysis, we adopt an NLP library named textacy\footnote{https://textacy.readthedocs.io/en/stable/} to extract semantic phrases, such as noun phrases (NP), verb phrases (VP), prepositional phrases (PP), and subject-verb object phrases (SVP).}
}

\textbf{Data Alignment of Gesture Data and Transcripts.}
Following the aforementioned steps for body keypoints extraction, we extract keypoints frame by frame from a presentation video. Then, we can obtain the timestamp for keypoints of a frame. As for speech transcripts, the transcripts with timestamps can be obtained by adopting automatic transcription techniques (e.g., Amazon Transcribe). Therefore, the keypoints and speech transcripts are naturally aligned based on timestamps.

\subsection{Task Analysis}
\textcolor{black}{Our goal is to develop a visualization system to assist professional presentation coaches in exploring and analyzing gesture usage of different presentations.} 
We followed a user-centered design process to derive analytical tasks and design our system iteratively. 
We worked closely with two experienced coaches (denoted as E1 and E2) from a presentation training company for about six months. 
Both E1 and E2 have at least five years of experience in the training of professional public speaking skills.
Their daily job is to offer professional communication courses to help trainees master presentation skills.
Gesture usage is one of the most important presentation skills they are focusing on.

\textcolor{black}{
To collect their requirements, we conducted independent interviews with the above target users (E1 and E2). 
During the interviews, we asked the coaches to 
(1) describe their general procedures and methods in the presentation training programs, especially for gesture usage;
(2) illustrate the challenges they meet in the training programs on gesture usage;
(3) clarify their desired system and functions for gesture analysis in presentation videos.}

\textcolor{black}{
Both coaches highlighted that gesture usage is important in presentations. 
They usually provided some guidelines and examples for students, 
since students are often unaware of their gestures used in presentations. 
Novice speakers are often too nervous and perform either few gestures or too many unconscious gestures.
Therefore, it is challenging to train students to get aware of their gestures.
To give their students concrete feedback, one of the methods is to record students' presentation videos. 
After that, coaches can manually review videos and offer students detailed suggestions. 
However, they are also facing issues in efficiently exploring gesture usage and generating quantitative comments. 
\textcolor{black}{E1 mentioned that it would be helpful to summarize gesture usage in presentation videos and provide example gestures for students. 
Further, E1 pointed out that it is useful to highlight repeated patterns, e.g., open and raise the hands up.} 
E2 emphasized the harmony between text/words and gestures, ``Make sure you are always saying with your body what your words are saying. For example, when I say large, my hands should express the meaning of large, and when I say tiny, my hands should show the meaning of tiny."
He expressed that it would be interesting to explore the connection between gestures and word phrases.
\textcolor{black}{For system design, both E1 and E2 mentioned that coaches often lack experience in working with complex visualization systems. Thus, they preferred to use intuitive visualization designs and a system that is easy to understand and use.}}

\textcolor{black}{Based on the coaches' feedback, we started to develop a visualization system. We iteratively refined our system by holding regular meetings with coaches and collecting their feedback.}
Here we summarize the derived analytical tasks:

\begin{itemize}
\item [\textbf{T1}]
\textbf{Obtain the spatial summary of gestures.} \textcolor{black}{Based on the interviews, a gesture summary can help coaches quickly explore gestures and identify examples.}
It is useful to obtain a summary of the spatial distribution of the hand movements, so that users can know where speakers tend to put their hands and what kind of gestures a speaker may employ. 
Such a spatial summary can also indicate speakers' gesture styles. 

\item [\textbf{T2}]
\textbf{Show the temporal evolution of gestures.} 
Since speakers can change gestures over time, it is necessary to explore the temporal evolution of gestures regarding the presentation content. 
A temporal summary of gestures makes users aware of how speakers move hands along time and gain deep insights into the temporal patterns of gesture usage.
Besides, users can know how often to use certain gestures.

\item [\textbf{T3}]
\textbf{\textcolor{black}{Explore the correlation between gestures and speech content.}}
\textcolor{black}{
According to our interviews with the coaches,
it is helpful for them to explore the correlation between gestures and speech content, e.g., whether the speakers use gestures that are harmonious with speech content. 
It is interesting to know what gestures are used to deliver certain content (content-based exploration).
Furthermore, it is useful to know what content different gestures tend to convey (gesture-based exploration).
}

\item [\textbf{T4}]
\textbf{\textcolor{black}{Find similar gestures used in presentations.}}
Research shows that speakers may unconsciously use similar or repetitive gestures of their own styles~\cite{tanveer2016automanner}. 
Our coaches expressed that it is helpful to explore similar or repetitive gestures, which can make speakers better aware of their own gestures.
Further, coaches can compare different similar gestures with different speech content, 
whereby understanding why speakers use such gestures and providing suggestions for improvements. 
For example, 
coaches could detect repetitive non-meaningful gestures made by speakers.

\item [\textbf{T5}]
\textbf{Enable interactive exploration of the presentation videos.} 
Coaches also need to check the original presentation video and the corresponding transcripts to confirm their findings of a speaker's gesture usage. 
Thus, it is necessary to provide coaches with interactive exploration of the original presentation videos.
\end{itemize}


\begin{figure*}[!t]
 \centering
 \includegraphics[width=1.88\columnwidth]{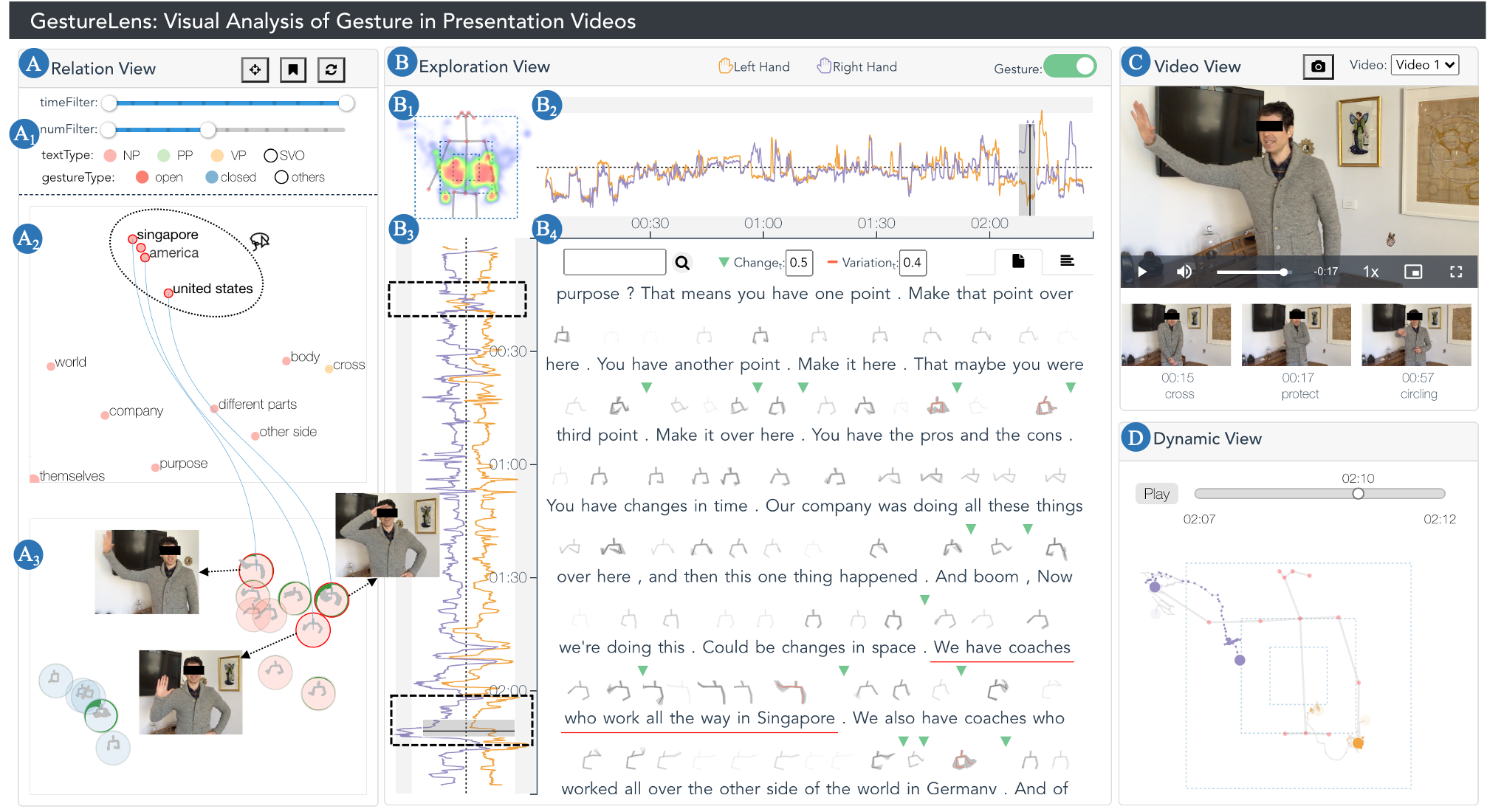}
  \caption{Our visualization system supports both gesture-based and content-based exploration of gesture usage.
  The {\gesLenContentview} (A) shows the connection between speech content ($\rm  A_1$) and gestures ($\rm  A_2$) by using a linked graph design.
  The {\gesLenMatrixview} (B) provides both spatial ($ \rm B_1$) and temporal ($\rm  B_2$ and $\rm  B_3$) summary of used gestures in the video, as well as transcripts and corresponding gestures ($\rm B_4$). 
  The {\gesLenVideoview} (C) provides the raw video. 
  The {\gesLenDynamicview} (D) shows gesture trajectories of selected gestures in the gesture space by using animation design.}
\label{fig:systemOverview}
 \vspace{-1em}
\end{figure*}

\section{System Overview}
In this section, we first describe the analytical pipeline of our system and then introduce each component of our system.

Fig.~\ref{fig:gesLenPipeline} shows our visualization system pipeline. 
Given presentation videos, we first conduct the data modeling phase (Section 3.1). 
After that, users can perform interactive visual exploration and analysis of gestures in presentation videos,
with gesture-based exploration and content-based exploration.

\textcolor{black}{
We implement our system based on the Vue.js front-end framework and the Flask back-end framework.
Our visual interface consists of four coordinated views. 
The {\gesLenContentview} (Fig.~\ref{fig:systemOverview}$\rm A$) shows the correlation between speech content and gestures (T3), as well as similar speech content and similar gestures (T4), which allows users to find what content gestures convey and what gestures are used to deliver the content. 
The {\gesLenMatrixview} presents a spatial summary of gesture usage (T1) with a heatmap in the gesture space (Fig.~\ref{fig:systemOverview}$\rm B_1$).
In addition, it visualizes
the temporal evolution of gestures (T2) along with the horizontal and vertical directions by showing
two perpendicular timelines (Fig.~\ref{fig:systemOverview}$\rm B_2$ and $\rm B_3$). Further, it annotates gesture glyphs on each word in the transcript, and highlights the dramatic changes in gestures (Fig.~\ref{fig:systemOverview}$\rm B_4$), which allows users to explore both gestures and speech content (T3). 
The {\gesLenVideoview} (Fig.~\ref{fig:systemOverview}C) provides detailed information on the speech content and gestures (T5). 
The {\gesLenDynamicview} (Fig.~\ref{fig:systemOverview}D) shows hand movement trajectories of selected gestures with animation.}


\section{Visualization Design}
In this section, we introduce the visual encoding of each view and the design alternatives we have considered, as well as the user interaction designs in the {\gesLen}.

\subsection{\gesLenContentviewCapital}
\textcolor{black}{
To explore what gestures are used in what speech content (T3),
we adopt a linked graph design to show the correlation between speech content and gestures.
}

\textit{Description}:
\textcolor{black}{
As shown in Fig.~\ref{fig:systemOverview}A, word phrases and gesture glyphs are projected into the middle 2D plane and the bottom 2D plane respectively by using the t-SNE algorithm. In this way, word phrases with similar meaning or similar gesture glyphs are close to each other in the corresponding plane. To reveal the relationship between gesture and its context, word phrases in the middle part and gesture glyphs at the bottom part are linked with lines. When users select word phrases or gesture glyphs of their interests, the corresponding gesture glyphs or word phrases are highlighted and linked with lines. For word phrases, according to our discussion with coaches, we extract the common word phrases of their interests, including noun phrases (NP), verb phrases (VP), prepositional phrases (PP), and subject-verb-object phrases (SVP) from the video transcript. 
Then, we convert these phrases into pre-trained Glove embeddings~\cite{pennington2014glove} for projection. As for gesture glyphs, we design a glyph for each gesture phase to better reveal patterns. As shown in Fig.~\ref{fig:systemOverview}$\rm A_3$, human stick figures are integrated into the centers of circles. The background colors encode three-type gestures, i.e., open, closed, and others. The green radial area chart along the circle represents the variation of gestures. A large green area indicates Large variation. 
To better configure this view, filtering configuration and legend are shown at the top (Fig.~\ref{fig:systemOverview}$\rm A_1$). Users are allowed to filter text phrases with a certain time range or occurrence number. Further, by clicking a legend type, the corresponding part can be shown or hidden. For example, as shown in Fig.~\ref{fig:systemOverview}$\rm A_1$, legend ``SVO (subject-verb-object)" and ``others" are not filled with colors, which indicates the corresponding types are filtered. Some other interactions are provided to facilitate exploration.
For example, when users click a word phrase or gesture glyph of interest, users can refer to the context of the word in the video, as well as the transcript area of the exploration view (Fig.~\ref{fig:systemOverview}$\rm B_4$). Further, bookmark interactions are provided to allow users to save the gestures of their interest, enabling future explorations.
}

\textcolor{black}{
\textit{Justification:}
We have considered other alternative designs. 
For example, we clustered similar gestures on the left side and placed the corresponding words on the right side using a word cloud design. 
Words in the word cloud are highlighted with different colors for different parts of speech (POS). 
\textcolor{black}{Our coaches confirmed the importance of explicitly exploring the connection between gestures and speech content.
However, they mentioned that the positions of words will be quite random in the word cloud design.
Also, they further suggested using some meaningful word phrases instead of using parts of speech (POS). 
Then we came up with the current linked graph design. 
Similar word phrases or similar gesture glyphs were close to each other.}
The corresponding word phrases and gesture glyphs can be connected with lines. 
Our coaches appreciated the linked graph design, which explicitly shows the connection between gestures and speech content.}

\subsection{\gesLenMatrixviewCapital}
To support intuitive and effective gesture-based and content-based explorations, we design an exploration view (Fig.~\ref{fig:systemOverview}B), which shows both spatial and temporal patterns of gestures, as well as corresponding content. 
Specifically, the exploration view contains three major components, i.e., a heatmap in the gesture space to reveal the spatial summary of gestures (T1), two mutually perpendicular timelines to uncover the temporal distribution of gestures (T2), and a transcript area to show the corresponding speech content and gestures (T3).

\textbf{Heatmap reveals spatial patterns.} 
As shown in Fig.~\ref{fig:systemOverview}$\rm B_1$, to reveal spatial patterns of gestures used in presentation videos, we layout a heatmap, a widely used technique for describing spatial patterns, over the gesture space. 
Three blue dashed rectangles are used to divide the gesture space into three areas, i.e., the center-center region, center region, and periphery region.
A human stick-figure skeleton is also shown in the gesture space, which can provide some context for the gesture space and heatmap. 
The heatmap in the gesture space can clearly reveal where a speaker tends to put his/her hands during a presentation. Our coaches like this design and confirm that it is easy to understand. Further, they think the heatmap can reveal speakers' presentation styles in terms of spatial patterns.

\textbf{Timelines describe temporal hand movements.} As shown in Fig.~\ref{fig:systemOverview}$\rm B_2$ and $\rm B_3$, two mutually perpendicular timelines are used to describe hand movements. These two timelines are aligned with the gesture space in the heatmap part (Fig.~\ref{fig:systemOverview}$\rm B_1$). The horizontal timeline (Fig.~\ref{fig:systemOverview}$\rm B_2$) describes the vertical position of two hands. The purple line indicates the right hand, while the orange line indicates the left hand. The horizontal dashed line is aligned with the vertical center of the gesture space, so we can observe the vertical position of two hands. At the bottom of this timeline, there is a click area for users to seek to the corresponding parts of the presentation videos. A vertical black line indicates the current time frame. Similarly, the vertical timeline (Fig.~\ref{fig:systemOverview}$\rm B_3$) is used to describe the horizontal position of two hands. The same visual encoding is applied in this timeline. From the vertical timeline, users can observe how users move their hands horizontally. For example, whether users use open hand gestures or closed hand gestures can be observed from this timeline. Besides, two timelines are linked together, i.e., when brushing one timeline, the corresponding place in the other timeline will be highlighted.

\textbf{Transcript area shows speech content.} 
As shown in Fig.~\ref{fig:systemOverview}$\rm B_4$, the speech content is explicitly shown here. 
To better reveal gestures used for speech content, we overlay human stick figures on top of the corresponding words. 
To be specific, gestures of the frames within a word are drawn on top of the word, which shows aggregate information of gestures for that word. 
\textcolor{black}{We further define \textit{spatial variation} to describe the variation of gestures within a word, and \textit{temporal change} to describe the change of gestures between two words. We first calculate the average gesture skeleton of each word, then we calculate the gesture skeleton variation within a word as spatial variation, while we calculate the gesture skeleton change between two words as temporal change.
We normalize both spatial variation and temporal change value to $[0\sim1]$. 
End users can customize the thresholds. 
As shown in Fig.~\ref{fig:systemOverview}$\rm B_4$, the change threshold and variation threshold are set to 0.5 and 0.4, respectively. 
As shown in the transcript area, high spatial variations are encoded with red strokes, and large temporal changes are encoded with green triangles.}

When users select interesting words, the corresponding time range will be highlighted in two timelines with a gray area. 
In reverse, when users select an interesting area in timelines, corresponding words will be highlighted.
Besides, users are allowed to search keywords by inputting a word at the top of the transcript area, and corresponding words will be highlighted with red underlines. 
To facilitate comparison with different gestures in the same word, a multiple-line mode showing corresponding sentences is provided.

\textit{Justification:}
\textcolor{black}{Before finally adopting this horizontal and vertical timeline-based design, we have considered other alternative designs.
Firstly, we drew trajectories in the gesture space to describe hand movements. However, it was not easy for users to track the temporal information. 
Then we considered using a timeline design to describe gestures. We first simply drew multiple lines with different styles to encode two hand movements (i.e., vertical movement and horizontal movement) in one timeline. However, with too many lines, it was not easy for users to analyze temporal patterns. Further, it was not easy for users to distinguish between different movements. After that, we considered a two-timeline design, one for describing the vertical movements and the other for describing the horizontal movements. However, this design could not distinguish the horizontal hand movements well. Therefore, we rotated one timeline to be vertical to better describe horizontal hand movements. After presenting these alternatives to our end users, they preferred it to the horizontal and vertical timeline-based design, which made it natural for them to observe temporal patterns. They also appreciated that we well-aligned these two timelines with the gesture space well, making it easy for them to conduct joint analysis with the spatial patterns.}

\subsection{\gesLenVideoviewCapital}
For video analysis, referring to the original video can sometimes provide a better explanation. Therefore, we embed the original video for exploration in the video view (Fig.~\ref{fig:systemOverview}C). After selecting a video of interest, the video is presented in this view. 
\textcolor{black}{Based on the coaches' feedback, the screenshot interaction is added to record some interesting moments. The corresponding information (i.e., time and word) is placed under screenshots, which can facilitate gesture usage exploration.} The video view is linked to other views. Users can refer to the video for detailed analysis (T5).

\subsection{\gesLenDynamicviewCapital}
\textcolor{black}{
The {\gesLenDynamicview}~(Fig.~\ref{fig:systemOverview}D) is designed to visualize gesture trajectories through animations, where users can play the animation and observe trajectories of the selected gesture. 
The gesture space is indicated with three blue dashed rectangles. 
A human stick-figure skeleton is presented to provide context for the gesture space. 
A gesture trajectory is encoded with a line and dots in the gesture space. Purple indicates the trajectory of the right hand and orange indicates the trajectory of the left hand. 
When users brush in the timelines of the exploration view to specify a time range, the corresponding gesture trajectories will be simultaneously updated in the {\gesLenDynamicview}. 
Users are allowed to play/pause the animation by pressing the button.}

\subsection{Interactions}
The four views in our system are linked and equipped with various user interactions, which provides strong visual analytic abilities. Here we summarize the interactions supported by our proposed system, {\gesLen}.

\textcolor{black}{
\textbf{Clicking.} 
In the {\gesLenVideoview}, users are allowed to seek different time frames by clicking the video timeline. Also, after clicking the timelines in the {\gesLenMatrixview}, the video then jumps to the corresponding time point.
When users click on words in the {\gesLenMatrixview}, word phrases and gesture glyphs in the {\gesLenContentview}, the video will be played from the corresponding time frame. Such clicking interactions allow users to refer to the raw video for further exploration.}

\textcolor{black}{
\textbf{Brushing.}
When users brush the horizontal and vertical timelines in the {\gesLenMatrixview}, the corresponding content will be highlighted in the transcript area.
When users brush the transcript area to select words of interest, the corresponding time range will be highlighted in the timelines. The brushing interaction is mainly used to select a focus area for exploration.}

\textcolor{black}{
\textbf{Searching.}
Our system enables users to search gestures by keywords.
When users input a keyword into the search area in the {\gesLenMatrixview}, the corresponding words will be found and highlighted with red underlines.}

\textcolor{black}{
\textbf{Configuration.}
Users are allowed to configure our system. For example, they can specify whether the gestures over words in the {\gesLenMatrixview} should be shown or not. Users can adjust the thresholds of the spatial variation and temporal change in the {\gesLenMatrixview}. Also, users are allowed to configure the filtering range in the {\gesLenContentview}.}


\section{Usage Scenario}
\label{section:usageScenarios}

We describe two usage scenarios to demonstrate the usefulness of {\gesLen} using a gesture practice video provided by our coaches and two TED talk videos.

\subsection{Analyzing open/closed gestures in a practice video}
\label{section:scenarioOne}
In this scenario, we describe how Lidia, a professional presentation coach, analyzes collected practice videos. Her goal is to obtain evidence about the speaker' performance on mastering the skills of gestures.
Also, the analysis results can provide some examples for her teaching later. 
Therefore, she explores the collected practice videos with {\gesLen}.

Lidia first selects a video in the video view. 
The video lasts about three minutes and records how a speaker practices gestures.
Then other views are updated correspondingly. 
To observe the spatial overview of the speaker's gestures (T1), Lidia shifts attention to the heatmap in the gesture space (Fig.~\ref{fig:systemOverview}$\rm B_1$). The large area of heatmap indicates that the speaker moves his hands intensely around the two sides within the center area of the gesture space. Also, the location of the less dense part of the heatmap indicates that the speaker sometimes stretches his hands outside the center area. It is normal for speakers to put their hands in front of their bodies most of the time. 
\textcolor{black}{According to Lidia's domain knowledge, she thinks that the speaker may tend to use closed and open gestures. Closed gestures refer to those gestures crossing arms or keeping hands close to the body, while open gestures refer to those gestures opening arms or moving hands far away from the body. These two types of gestures are widely used in presentations. Lidia feels interested in whether the speaker uses appropriate closed and open gestures.} Therefore, Lidia need to find out when the speaker uses closed gestures and open gestures. She further observes the two mutually perpendicular timelines (Fig.~\ref{fig:systemOverview}$\rm B_2$ and $\rm B_3$), where she can obtain a temporal summary of gestures (T2).
The purple and orange lines indicate the movements of the right hand and left hand, respectively. 
From the horizontal timeline, the speaker first puts his hands below the dashed centerline, then moves his hands up to be around the dashed centerline. 
At the end, the speaker makes large vertical movements in the vertical direction. 
From the vertical timeline, the speaker sometimes moves both hands together, both hands to one side, and moves both hands away. 
This information confirms that the speaker uses some closed and open gestures in some moments.

\textcolor{black}{Then, she decides to explore details by examining interesting moments about closed and open gestures (T2).}
For example, Lidia examines a moment when the purple line and orange line are close, as shown in the first black dashed rectangle in Fig.~\ref{fig:systemOverview}$\rm B_3$. Lidia finds that the corresponding speech content is ``Locked in, they cross their body". After watching the raw video (T5), Lidia realizes that the speaker moves both hands close to each other to express what he is saying. \textcolor{black}{Lidia believes that it is a good example of using closed gestures.}
Further, Lidia examines a moment of the large movements, where the purple line and orange line are far away, as shown in the second black dashed rectangle in Fig.~\ref{fig:systemOverview}$\rm B_3$. The corresponding speech content is ``...who work all the way in Singapore ...., who work all over the other sides of the world in Germany". Lidia further checks the gestures used for this content in the video view and dynamic view. As shown in Fig.~\ref{fig:systemOverview}$\rm C$ and $D$, \textcolor{black}{she finds that the speaker opens his hands to express different locations and long distances, which is a good example of using open gestures.}

\begin{figure}[!t]
  \setlength{\belowcaptionskip}{-10pt}
  \centering
  \includegraphics[width=1\columnwidth]{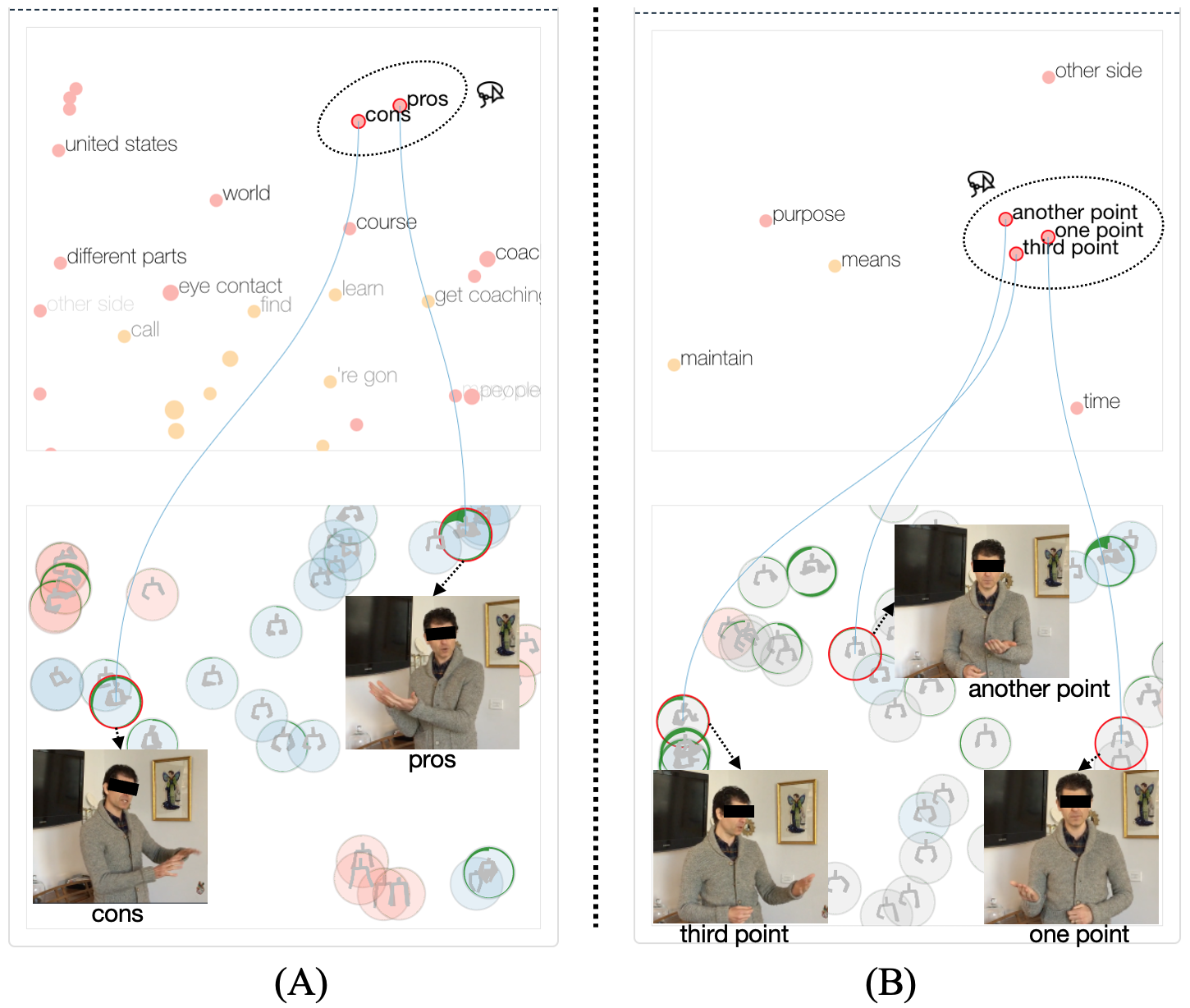}
  \vspace{-5mm}
  \caption{Exploring gestures for a set of words. (A) The speaker moves his hands to the right or left-hand sides to express two viewpoints, i.e., ``pros" and ``cons". (B) The speaker moves his hands from the right-hand side, to the middle and left-hand side to enumerate different viewpoints, i.e., ``one point", ``another point" and ``third point".}
  \label{fig:gesLenCaseOneRelationSimilarWords}
\end{figure}

\begin{figure}[!t]
  \setlength{\belowcaptionskip}{-10pt}
  \centering
  \includegraphics[width=1\columnwidth]{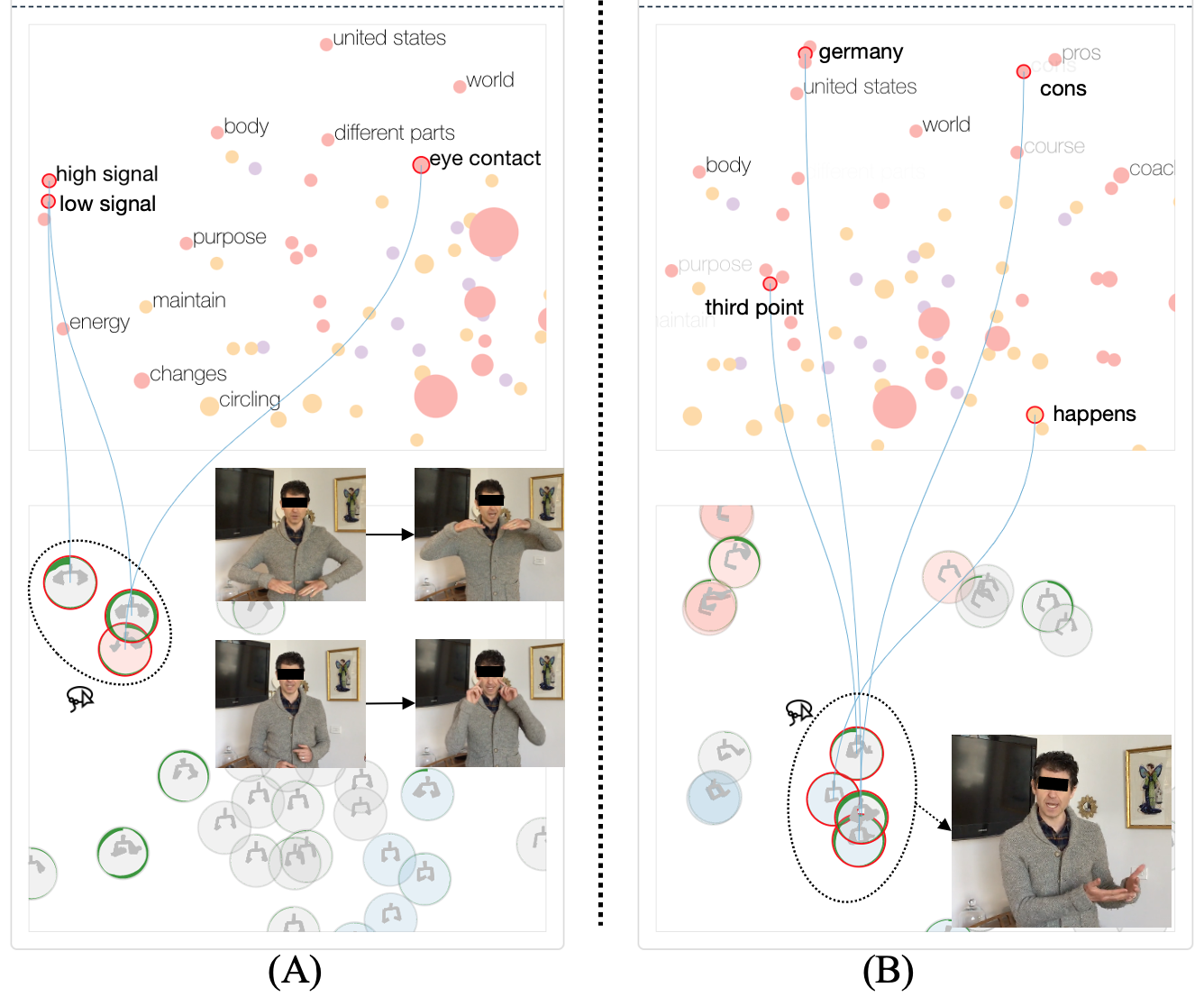}
  \vspace{-5mm}
  \caption{Exploring speech content for a group of similar gestures. (A) The speaker moves his hands from the bottom to top to express the meanings of ``low signal" to ``high signal" and ``eye contact". (B) The speaker moves his hands to the left-hand side to express different ideas, such as locations and viewpoints.}
  \label{fig:gesLenCaseOneRelationSimilarGestures}
\end{figure}

\textcolor{black}{To further explore the correlation between gesture and speech content (T3), Lidia shifts her attention to the relation view.} As shown in Fig.~\ref{fig:systemOverview}$\rm A_2$, to find out gestures used to express location, she selects several locations, e.g., America, Germany and Singapore in the noun phrases. Corresponding gestures are highlighted at the bottom and linked with the selected word phrases. By examining each gesture glyph, Lidia finds the speakers tend to use open gestures to express locations that are far away from each other, such as moving his hands to the right side or left side.
As shown in Fig.~\ref{fig:gesLenCaseOneRelationSimilarWords}A, to find out gestures that are used to express opposite viewpoints, Lidia selects two words, ``pros" and ``cons". By examining corresponding gesture glyphs, she notices that the speaker tends to use two opposite directions (the right-hand and left-hand sides) to express opposite viewpoints. Similarly, as shown in Fig.~\ref{fig:gesLenCaseOneRelationSimilarWords}B, \textcolor{black}{Lidia finds the speaker tends to move his hands to different directions (the right-hand side, middle, and left-hand side) to enumerate different viewpoints. Lidia thinks that it is quite common for speakers to use deictic gestures to draw attention to different objects and representational gestures (e.g., metaphoric gestures) to refer to objects or events.}
\textcolor{black}{Also, Lidia wants to explore speech content used for similar gestures (T4). Lidia selects several similar gestures, the corresponding word phrases are highlighted and linked (Fig.~\ref{fig:gesLenCaseOneRelationSimilarGestures}).} After checking the details, as shown in Fig.~\ref{fig:gesLenCaseOneRelationSimilarGestures}A, she finds that the speaker moving his hands from the bottom to top to express the meanings of ``low signal" to ``high signal". Similarly, the speaker emphasizes the meaning of ``eye contact" by moving his hands from the bottom to top for pointing his eyes. What is more, as shown in Fig.~\ref{fig:gesLenCaseOneRelationSimilarGestures}B, the speaker moves his hands to the left-hand side to enumerate different locations (e.g., ``Germany") or viewpoints (e.g., ``third point"). Meanwhile, Lidia is curious about the word ``happens" since this word is not related to enumeration at first glance, then she checks the details and finds that ``happens" corresponds to the context ``what happens next", which is also used to introduce a subsequent viewpoint.

In summary, after analyzing the video with {\gesLen}, \textcolor{black}{Lidia confidently concludes that the speaker masters gesture usage quite well. The speaker can use appropriate gestures for different speech content, especially the closed gestures and open gestures. Also, she thinks that this video demonstrates a good usage of metaphoric gestures, i.e., using hand movements to represent abstract ideas or concepts, which is a good example for her teaching}.

\begin{figure*}[!t]
  \setlength{\belowcaptionskip}{-5pt}
  \centering
  \includegraphics[width=2\columnwidth]{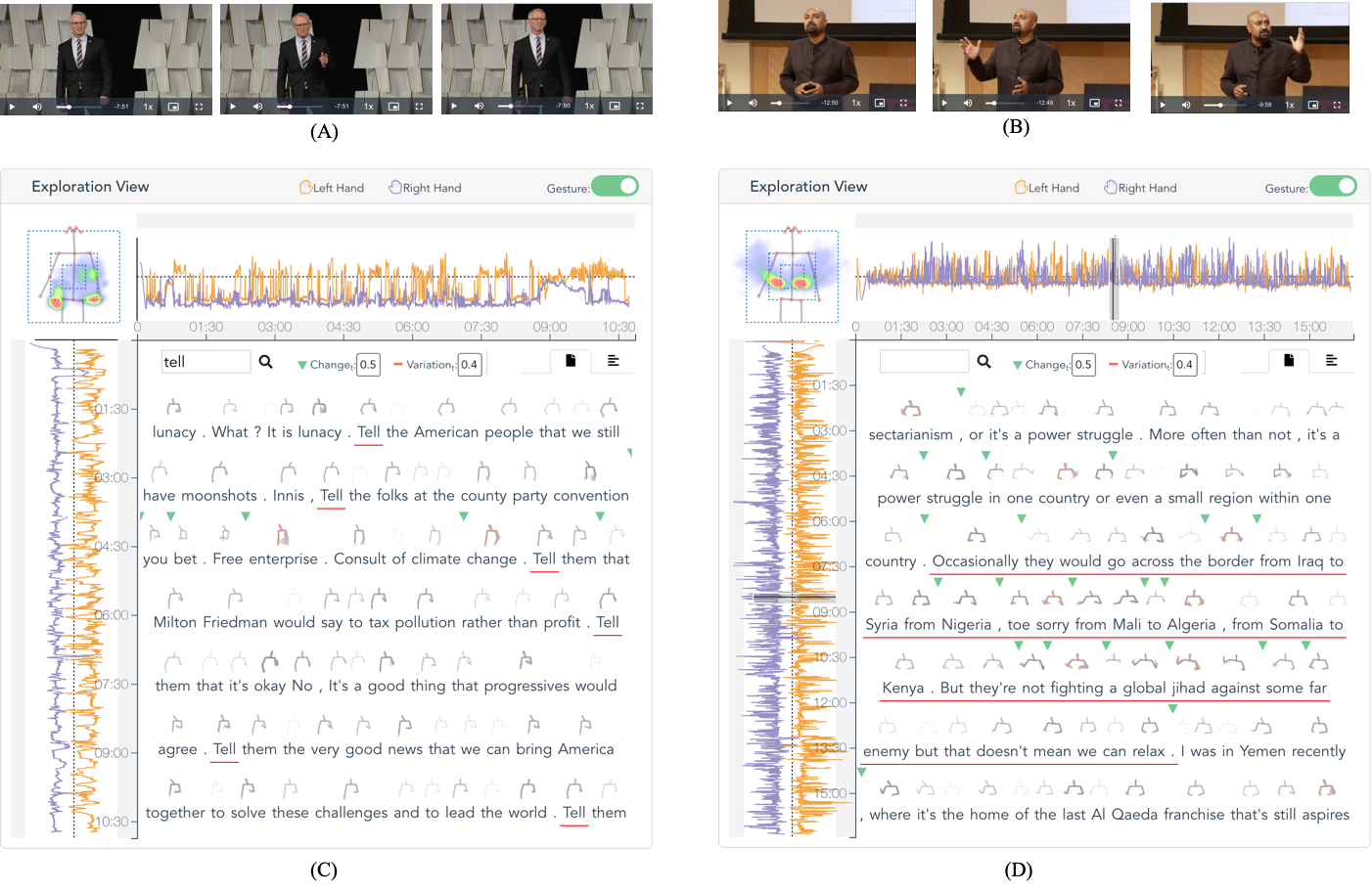}
  \vspace{-3mm}
  \caption{Beat gestures in two TED talk videos. (A) Screenshots show raising the left hand and putting the left hand down in the first TED video. (B) Screenshots show moving hands to the left and right sides in the second TED video. (C) The exploration view shows few hand movements of the right hand with purple lines and intense hand movements of the left hand with orange lines in the first TED video. (D) The exploration view shows the symmetrical movements of the left and right hands with orange and purple lines respectively in the second TED video.}
  \label{fig:gesLenCaseTwo}
\end{figure*}

\subsection{Exploring beat gestures of TED talk videos}
\label{section:scenarioTwo}
In this scenario, we describe another professional presentation coach, Kevin, who usually uses TED talk videos as examples in his presentation training classes.
During his teaching, he needs to analyze TED videos and show some examples to his students.
In this scenario, he explores and compares two TED talk videos, one is named ``American bipartisan politics can be saved — here's how", which is about eleven minutes long; and the other video named ``Why global jihad is losing" lasts seventeen minutes long.

After Kevin selects the first video, the corresponding views are updated.
He can observe the spatial and temporal distribution of the gestures (T1-2) from the exploration view (Fig.~\ref{fig:gesLenCaseTwo}C). 
The heatmap in the gesture space indicates the speaker mainly puts his hands in the low position. Sometimes, he raises his left hand up.
This pattern can further be examined by the horizontal timeline, where the purple line mainly stays in the low position and the orange line fluctuates all the time. 
As for the vertical timeline, his right hand (purple line) stays almost in the same horizontal place, while his left hand (orange line) shows some movements, e.g. moving to the center and moving back to the left side. 
After referring to the original video, Kevin notices that the speaker's right hand is carrying something so that it does not move too much at all. 
While the speaker sometimes raises his left hand when he wants to emphasize some words. 
In addition, the speaker tends to use parallel sentence structures. 
For example, as shown in the transcript area of Fig.~\ref{fig:gesLenCaseTwo}C, the speaker uses ``tell" many times, which are highlighted with red underlines. The speaker repetitively raises his left hand and quickly puts it down to emphasize the word ``tell".
The corresponding gesture is shown in Fig.~\ref{fig:gesLenCaseTwo}A.

To explore another style of gestures, Kevin selects the second video.
As shown in Fig.~\ref{fig:gesLenCaseTwo}D, the spatial and temporal patterns are quite different from the first video (T1-2). The heatmap in the gesture space indicates that the speaker tends to use his hands symmetrically and puts his hands in front of his stomach most of the time.
Also, the speaker sometimes moves his hands to the left or right side. 
This pattern is further examined in the two mutually perpendicular timelines. His hands mainly move around the above dashed centerline in the horizontal timeline, which is indicated by the fluctuating purple and orange lines, as shown in Fig.~\ref{fig:gesLenCaseTwo}D.
Also, his right hand (purple line) and left hand (orange line) move around the left side and right side respectively in the vertical timeline. After referring to the raw video (T5), Kevin finds the corresponding gesture (Fig.~\ref{fig:gesLenCaseTwo}B).
Compared with the speaker in the first video, the hand movements of this speaker are more intense. 
As shown in the transcript area of Fig.~\ref{fig:gesLenCaseTwo}D, there are more green triangles and red strokes, which further indicates that the speaker is excited and tends to move his hands more intensely. 

Overall, \textcolor{black}{Kevin finds that speakers in both videos mainly used beat gestures, i.e., the rhythmic gestures following alongside the natural stress patterns of speech.} Although most of the time, the gestures used in these two videos do not have a clear meaning, these gestures demonstrate the rhythm of the speakers. \textcolor{black}{According to his domain knowledge, Kevin confirms that these two videos are good examples for demonstrating beat gestures with different styles, i.e., moving hands up and down to emphasize some words in parallel sentence structures and moving hands to left and right sides symmetrically to emphasize another set of words. 
Basically, the beat gestures have been employed by the speakers to enhance the message delivery in their presentations.}


\section{Expert Interview}
To further evaluate the usefulness and effectiveness of {\gesLen}, we conducted semi-structured interviews with our aforementioned collaborating domain experts (E1 and E2) and two other domain experts (E3 and E4) who are not involved in the design process. E3 and E4 have three and five years of experience in the training of professional public speaking skills, respectively.
Four experts have basic knowledge about simple visualization techniques, such as bar charts. 
Each interview lasted about an hour.

During each interview, we first briefly introduced the interface and functions of {\gesLen} and then presented two usage scenarios described in Section 6 to our four domain experts. All of them appreciated the two usage scenarios and thought that the two usage scenarios clearly demonstrated the functions and effectiveness of {\gesLen}.
Then considering the analytical tasks of exploring gestures in presentation videos, we designed the following tasks to guide their open-ended explorations:
\begin{compactitem}
\item Observe and describe the spatial and temporal summary of gestures used in the selected video.
\item Inspect some moments of interest and find out corresponding gestures and speech content.
\item \textcolor{black}{Explore what gestures are used for some word phrases of interest.}
\item Judge whether the speaker uses appropriate gestures for the corresponding speech content.
\end{compactitem}

After the experts finished these tasks, we collected their feedback about the system.
In general, their feedback was positive and they appreciated the system of exploring gesture usage in presentation videos with visualization techniques. Here we summarized their feedback as follows:

\textcolor{black}{
\textbf{System Designs.}
Both E1 and E2 appreciated the visual designs, especially the exploration view that can help them explore gestures in presentation videos. 
E1 thought that they could use the four views to accomplish the identified analytical tasks quite well.
He said~\textit{``It is smart to decompose dynamic movements into two mutually perpendicular timelines, which can reveal some detailed information I usually ignore. For example, it is hard for me to pay attention to all the closed gestures in a presentation."} 
E2 mentioned that~\textit{``The system interface has improved a lot after several iterations. Currently, the system can help us analyze the hand movements of speakers. It is nice to have a graph (representation) of gestures when providing feedback."}
Although E3 and E4 were not involved in the design process, they can quickly understand the designs after a brief introduction.
E3 felt interested in exploring the connection between gestures and speech content.
He said~\textit{``I like the idea of matching gestures and movements with the specific words, which can show evidence for my students."} 
E4 appreciated the interactions we provided. 
He mentioned that~\textit{``The interactions are quite useful, especially allowing users to refer to the origin video."}}

\textcolor{black}{
\textbf{Usability and Improvements.}
All the experts agreed that the exploration view, video view and dynamic view are easy to understand and use. 
\textcolor{black}{For the relation view, E3 and E4 stated that the relation view is easy to understand but takes a little effort to master.}
E3 considered this view as an advanced function when exploring the connection between gestures and speech content.
E3 mentioned that~\textit{``Other views can provide an overview of gestures and their context for me. This view can provide in-depth exploration."} 
Both E3 and E4 suggested adding more functions to analyze gestures in presentations. 
For example, E3 mentioned~\textit{``It will be useful to analyze fingers in presentation videos, as well as integrating other information, such as facial expressions and audio features."} 
E4 further suggested recommending gestures for different speech content instantly. 
He said,~\textit{``We can help speakers tailor a better presentation if we advise appropriate gestures based on the speech content and tell them how memorable are the gestures they are using."}}


\section{Discussion And Limitations}
In this section, we discuss the lessons learned and reflections when developing {\gesLen}.

\textbf{Lessons Learned for Gesture Usage.} 
Here we present two lessons learned for gesture usage when closely working with coaches. 
First, video-based gesture training is effective. Domain experts confirm that analyzing presentation videos could provide clear evidence for improving gesture usage. By exploring presentation videos, coaches could better analyze speakers' gestures and find some good gesture usage examples. Then speakers could effectively practice their gestures based on coaches' feedback.
Second, gestures should facilitate better content delivery without distracting audiences. Here we describe two concrete gesture styles, i.e.,~\textit{metaphoric gestures} and~\textit{beat gestures}. 
~\textit{Metaphoric gestures} are one of the most widely used styles in presentations, where speakers use hand movements to represent abstract ideas or concepts. 
For example, as shown in the first scenario (Section 6.1), the speaker opens his hands (open gestures) to express different locations and long distances. Also, the speaker moves his hands to different directions (e.g., the right-hand side, middle and left-hand side) to enumerate different viewpoints.
~\textit{Beat gestures} are another important gesture wildly used. But many people may move their hands casually and it can even distract audiences. The two speakers in the second scenario (Section 6.2) provide good examples for using~\textit{beat gestures} to emphasize some words, which demonstrates that gestures should be well correlated with the speech content.

\textcolor{black}{
\textbf{Reflection on Visual Design.}
Here we present two reflections on visual design during developing {\gesLen}.}
First, it is important to strike a good balance between the intuitiveness of the visual designs and the amount of encoded information when designing the visualizations, especially for users without much background in visualization (e.g., the coaches in this paper).
At the initial stage of our design, we tend to employ some relatively complex visualizations to encode more information. However, these designs were rejected by the coaches, as they feel overwhelmed by the information encoded in those designs.
Second, it is not easy to achieve a detailed and comprehensive exploration of gestures due to their intrinsic complexity.
The complexity of gestures originates from their spatial distribution (gesture space) and temporal evolution (dynamic movements), as well as the correlation between gesture usage and the speech content.
In this study, we have explicitly handled this from two perspectives: the exploration view (Fig.~\ref{fig:systemOverview}$\rm B$) decomposes gestures to temporal and spatial dimensions for an easy representation and exploration of gesture details, and the relation view (Fig.~\ref{fig:systemOverview}$\rm A$) visualizes the overall correlation between gestures and the major topics of speech content.
Such visual designs are appreciated by our target users, as demonstrated in the above expert interviews. 

\textbf{Generalizability.}
{\gesLen} is initially designed for analyzing the gestures of a speaker in presentation videos, but it can also be generalized to analyze other related human behaviors in other scenarios, \textcolor{black}{such as analyzing gesture usage in sign language, analyzing gesture usage of a music conductor, and analyzing body postures of dancers. For the detailed visualization designs proposed in {\gesLen}, the design of two mutually perpendicular timelines can be applied to analyze the spatial distribution and temporal evolution of gestures used in sign language. The human stick-figure glyphs can be applied to reveal and compare different postures of dancers following kinds of music.}

We also identify several limitations that need further research in future work, including gesture analysis, accuracy, scalability and evaluation.

\textbf{Gesture Analysis.}
In this paper, we visualize the spatial distribution and temporal evolution of gestures which enables coaches to explore how gestures correlate with speech content. Here are some discussions on our approaches. First, for the research scope, we are aware that we mainly focus on the hand movements in the upper body. As suggested by coaches, it would be better to analyze more details, such as hand-shape and finger gestures. Based on the current research, we can further analyze hand shapes and fingers in the future.
Second, for the gesture data extraction, we are aware that if a speaker largely shifts his stance, one camera cannot fully capture his gestures. We need multiple cameras to capture the presenter's gestures from different camera shots, angles, and movements. Then better normalization and integration methods are needed to conduct gesture analysis, which is left as future work.
Third, {\gesLen} mainly analyzes the relationship between gestures and speech content. It would be also interesting to explore the correlation between gestures and other multi-modal features, such as facial expressions and audio features.

\textbf{Accuracy.}
We extract body keypoints of gestures by using a widely-used library, which achieves an accuracy of around $75.6\%$ mean Average Precision (mAP) on the MPII dataset~\cite{Cao2019openpose}. As for speech transcription, we adopt a wildly-used API named Amazon Transcribe, which achieves an accuracy of $94\%$\footnote{https://cloudcompiled.com/blog/transcription-api-comparison/}. Although the extracted data could not achieve an accuracy of $100\%$, it does not affect the analytical process too much. Our coaches are satisfied with the performance of body keypoint detection and speech transcription. In the future, we can consider more advanced approaches to improve the data extraction and encode the uncertainty to inform users of the potential inaccuracy.

\textbf{Scalability.}
{\gesLen} is mainly applied to analyze the gesture usage of presentation videos lasting about 10 minutes.
However, with the increasing of the presentation length and complexity of gesture usage, {\gesLen} may have some scalability issues. 
Due to the intrinsic dynamic nature of gestures, a gesture often involves a number of frames. When a presentation contains rich gestures, visual clutter issues may arise in the timelines of the {\gesLenMatrixview} and the {\gesLenDynamicview}, 
due to the limited screen size.
\textcolor{black}{Such kinds of visual clutter issues can be handled by some straightforward strategies. For example, integrating some automated methods to extract some interesting parts and filter out some insignificant frames.}

\textcolor{black}{
\textbf{Evaluation.}
We have presented two usage scenarios and conducted expert interviews with four professional presentation coaches, which provides support for the usefulness and effectiveness of {\gesLen}. However, we acknowledge that it is better to recruit more professional coaches to conduct human-subject studies, which is left as future work.
}


\section{Conclusion and Future Work}
In this paper, we propose an interactive visual analytics system, {\gesLen}, to facilitate gesture exploration in presentation videos. The system provides both a spatial and temporal overview of gestures, as well as a detailed analysis of gestures.
Specifically,
we extend and integrate some well-established visualization techniques to intuitive visualization designs, e.g., two mutually perpendicular timelines to reveal the temporal distribution of gestures.
Two usage scenarios and interviews with domain experts are conducted to demonstrate the effectiveness of our system. 

In the future, we plan to improve the system usability by adding more functions, such as gesture style comparison and gesture recommendation. Also, we would like to consider multi-modal features (e.g., facial expressions and audio) and incorporate more advanced data mining techniques to enhance gesture analysis. Furthermore, we will conduct a long-term study with more domain experts to further evaluate the usability and effectiveness of {\gesLen}.



%

%
%

\ifCLASSOPTIONcompsoc
  \section*{Acknowledgments}
\else
  \section*{Acknowledgment}
\fi
\textcolor{black}{
We would like to thank our industry collaborator for offering valuable resources. We also thank our domain experts and the anonymous reviewers for their insightful comments. This work is partially supported by the 100 Talents Program of Sun Yat-sen University and a grant from ITF PRP (Project No. PRP/001/21FX).}



\ifCLASSOPTIONcaptionsoff
  \newpage
\fi



\bibliographystyle{IEEEtran}
\bibliography{main.bib}
\newpage
\balance
\begin{IEEEbiography}[{\includegraphics[width=1in,height=1.25in,clip,keepaspectratio]{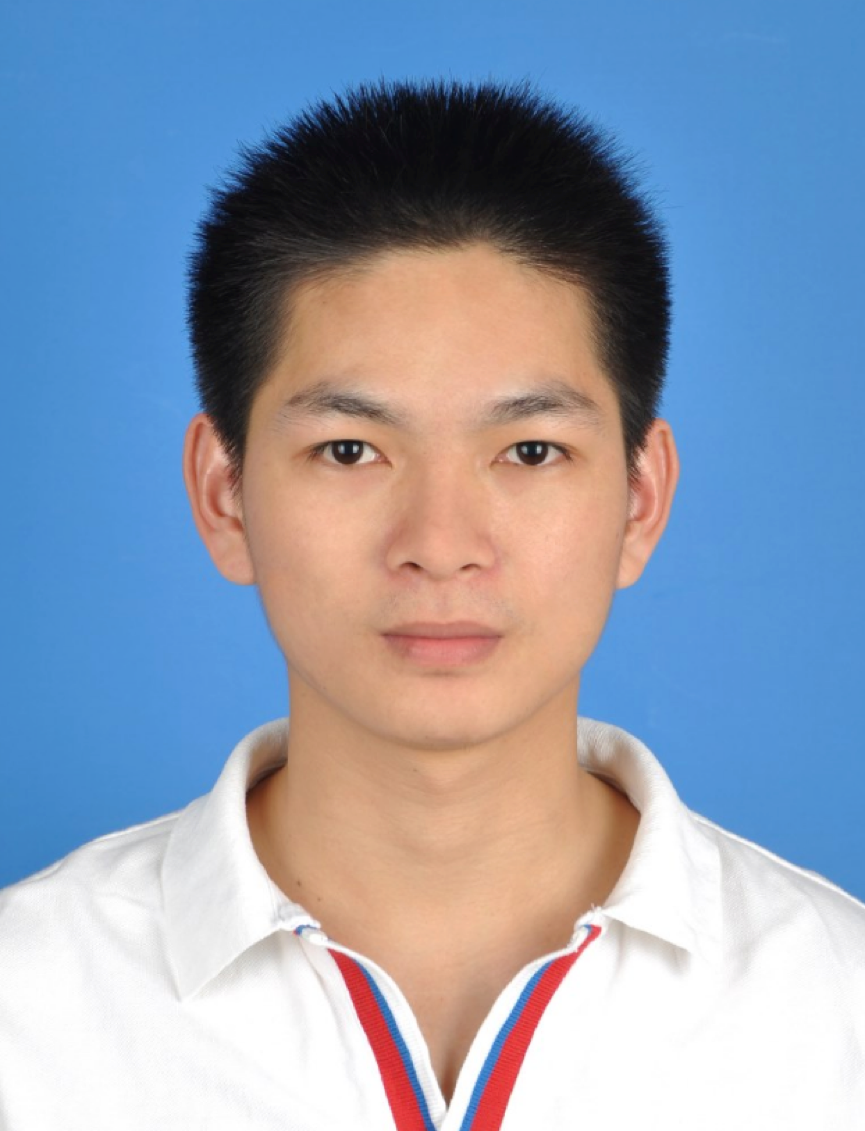}}]{Haipeng Zeng}
is currently an assistant professor in School of Intelligent Systems Engineering at Sun Yat-sen University (SYSU). He obtained a B.S. in Mathematics from Sun Yat-Sen University and a Ph.D. in Computer Science from the Hong Kong University of Science and Technology. His research interests include data visualization, visual analytics, machine learning and intelligent transportation. For more details, please refer to \url{http://www.zenghp.org/}.
\end{IEEEbiography}
 \vspace{-5mm}
 
\begin{IEEEbiography}[{\includegraphics[width=1in,height=1.5in,clip,keepaspectratio]{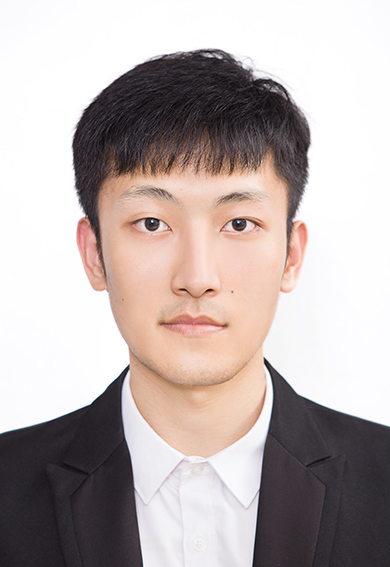}}]{Xingbo Wang}
is a Ph.D. candidate in the Department of Computer Science and Engineering at the Hong Kong University of Science and Technology (HKUST). He obtained a B.E. degree from Wuhan University, China in 2018.
His research interests include multimedia visualization, interactive machine learning for natural language processing (NLP).
For more details, please refer to \url{https://andy-xingbowang.com/}.
\end{IEEEbiography}
 \vspace{-5mm}

\begin{IEEEbiography}[{\includegraphics[width=1in,height=1.5in,clip,keepaspectratio]{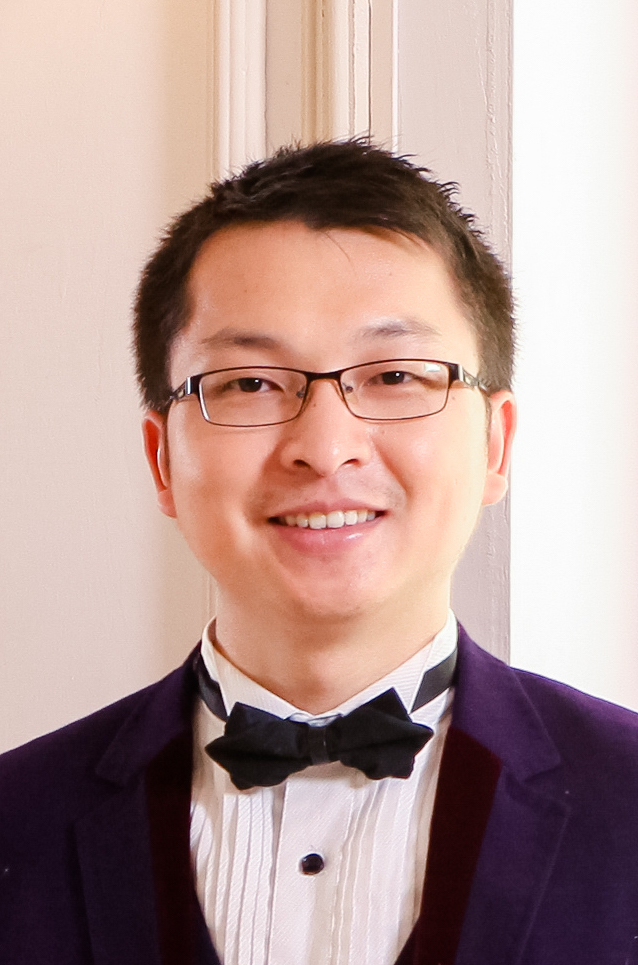}}]{Yong Wang} is currently an assistant professor in School of Computing and Information Systems at Singapore Management University. His research interests include data visualization, visual analytics and explainable machine learning.
He obtained his Ph.D. in Computer Science from Hong Kong University of Science and Technology in 2018. He received his B.E. and M.E. from Harbin Institute of Technology and Huazhong University of Science and Technology, respectively. For more details, please refer to \url{http://yong-wang.org}.
\end{IEEEbiography}
 \vspace{-5mm}
 
\begin{IEEEbiography}[{\includegraphics[width=1in,height=1.25in,clip,keepaspectratio]{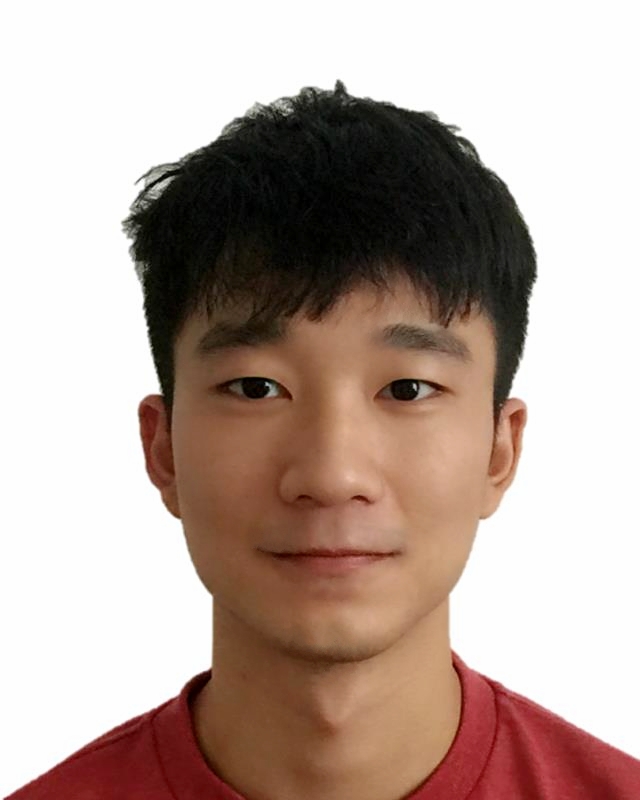}}]{Aoyu Wu} is a Ph.D. candidate in the Department of Computer Science and Engineering at the Hong Kong University of Science and Technology (HKUST). He received his B.E. and M.E. degrees from HKUST. His research interests include data visualization and human-computer interaction.
For more details, please refer to \url{https://wowjyu.github.io/.}
\end{IEEEbiography}
 \vspace{-5mm}

\begin{IEEEbiography}[{\includegraphics[width=1in,height=1.25in,clip,keepaspectratio]{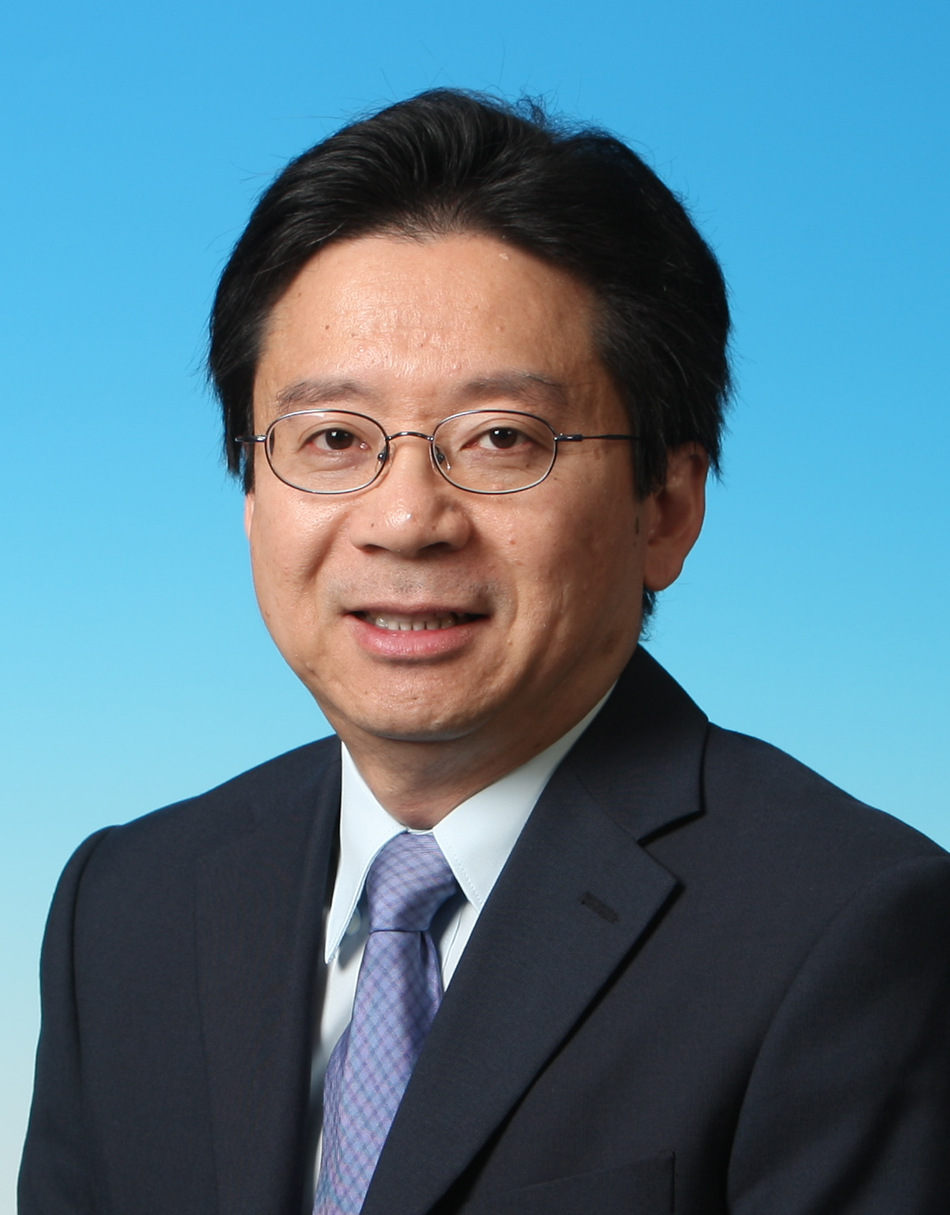}}]{Ting-Chuen Pong}
is a professor of the Department of Computer Science and Engineering at the Hong Kong University of Science and Technology (HKUST). He received his Ph.D. in Computer Science from Virginia Polytechnic Institute and State University in 1984. His research interests include computer vision, multimedia computing and IT in Education. For more information, please visit \url{http://www.cse.ust.hk/faculty/tcpong/}.
\end{IEEEbiography}
 \vspace{-5mm}

\begin{IEEEbiography}[{\includegraphics[width=1in,height=1.5in,clip,keepaspectratio]{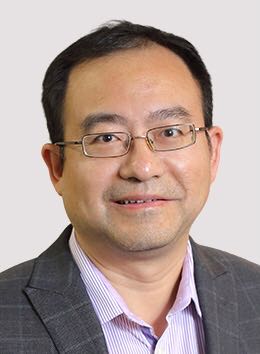}}]{Huamin Qu}
is a professor in the Department of Computer Science and Engineering (CSE) at the Hong Kong University of Science and Technology (HKUST) and also the director of the interdisciplinary program office (IPO) of HKUST. He obtained a BS in Mathematics from Xi'an Jiaotong University, China, an MS and a PhD in Computer Science from the Stony Brook University. His main research interests are in visualization and human-computer interaction, with focuses on urban informatics, social network analysis, E-learning, text visualization, and explainable artificial intelligence (XAI). For more information, please visit \url{http://huamin.org/}.
\end{IEEEbiography}
 \vspace{-5mm}

\end{document}